\documentclass[twocolumn,english,showpacs]{revtex4}
\usepackage{babel,amsmath,amssymb,dcolumn}
\usepackage[dvips]{graphics}
\def\imo{i}
\def\agt{\mathrel{\raise.3ex\hbox{$>$}\mkern-14mu\lower0.6ex\hbox{$\sim$}}}
\def\alt{\mathrel{\raise.3ex\hbox{$<$}\mkern-14mu\lower0.6ex\hbox{$\sim$}}}

\def\beq{\begin{equation}}
\def\eeq{\end{equation}}
\def\bsubeq{\begin{subequations}}
\def\esubeq{\end{subequations}}

\begin{document}
\title{Quasinormal modes of black holes immersed in a strong magnetic field}
\author{R. A. Konoplya}
\email{konoplya@fma.if.usp.br}
\author{R. D. B. Fontana}
\email{rodrigof@fma.if.usp.br}
\affiliation{Instituto de F\'{\i}sica, Universidade de S\~{a}o Paulo \\
C.P. 66318, 05315-970, S\~{a}o Paulo-SP, Brazil}

\begin{abstract}
We found quasinormal modes, both in time and frequency domains, of
the Ernst black holes, that is neutral black holes immersed in an
external magnetic field. The Ernst solution reduces to the
Schwarzschild solution, when the magnetic field vanishes. It is
found that the quasinormal spectrum for massless scalar field in
the vicinity of the magnetized black holes acquires an effective
"mass" $\mu = 2 B m$, where $m$ is the azimuthal number and $B$ is
parameter describing the magnetic field. We shall show that in the
presence of a magnetic field quasinormal modes are longer lived
and have larger oscillation frequencies. The perturbations of
higher dimensional magnetized black holes by Ortaggio and of
magnetized dilaton black holes by Radu are considered.
\end{abstract}

\maketitle
%%%%%%%%%%%%%%%%%%%%%%%%%%%%%%%%%%%%%%%%%%%%%%%%%%%%%%%%%%%%%%%%%%%%%%%%%%%%%%%
%\section{Introduction}
%%%%%%%%%%%%%%%%%%%%%%%%%%%%%%%%%%%%%%%%%%%%%%%%%%%%%%%%%%%%%%%%%%%%%%%%%%%%%%%
It is well known that supermassive black holes in centres of
galaxies are immersed in a strong magnetic field. Interaction of a
black hole and a magnetic field can happen in a lot of physical
situations: when an accretion disk or other matter distribution
around black hole is charged, when taking into consideration
galactic and intergalactic magnetic fields, and, possibly, if
mini-black holes are created in particle collisions in the
brane-world scenarios. Let us note that in Large Hadron Colliders
in the region of particle collisions, the huge accelerating
magnetic field is assumed to be screened, yet this does not
exclude possibility of interaction of strong magnetic fields and
mini-black holes in a great variety of high energy processes, when
quantum gravity states are excited.

In addition to highly motivated astrophysical interest to magnetic
fields around black holes \cite{magnetic-astro}, these fields are
important also as a background field testing the black hole
geometry. Thus, when perturbed, the magnetic field undergoes
characteristic damped oscillations, quasinormal modes, which could
be observed in experiments. Quasinormal modes of black holes has
gained considerable attention recent few years also because of
their applications in string theory through the AdS/CFT
correspondence. As nowadays there are a great number of papers on
this subject, we refer the reader to the reviews
\cite{Kokkotas-99} and a few papers \cite{QNMrecent} where a lot of references
to the recent research of quasinormal modes can be found.

In the present paper we consider the massless scalar field
perturbations around the Ernst black hole \cite{Ernst}, a black
hole immersed in an external magnetic field, and around its higher
dimensional \cite{Ortaggio:2004kr} and dilatonic
\cite{Agop:2005np} generalizations.
The Ernst metric is the exact solution of the Einstein-Maxwell
equations. Now, the properties of the Ernst metric are well
studied, since the discovery of the Ernst solution
\cite{Ernst-properties}, and, different generalizations of the
Ernst solution are obtained
\cite{Gorbatsievich:1995jz}. In the present paper, we shall find the
quasinormal modes of the Ernst black holes both in time and
frequency domain and shall show how magnetic field affect the
quasinormal spectrum.

The Ernst metric in four dimensions has the form
\cite{Ernst},
$$
d s^{2} = \Lambda^{2} \left(
\left(1- \frac{2 M}{r} \right) d t^{2} - \left(1- \frac{2 M}{r} \right)^{-1} d
r^{2} -r^{2} d \theta^{2}
\right),
$$
\begin{equation}
 - \frac{r^{2} \sin^{2} \theta}{\Lambda^{2} } d \phi^{2},
\end{equation}
where the external magnetic field is determined by the parameter
$B$
\begin{equation}
\Lambda = 1 + B^{2} r^{2} \sin^{2} \theta.
\end{equation}
The vector potential for the magnetic field is given by the
formula:
\begin{equation}
A_{\mu} d x ^{\mu} = - \frac{B r^{2} \sin^{2} \theta} {\Lambda} d
\phi.
\end{equation}
As a magnetic field is assumed to exist everywhere in space, the
above metric is not asymptotically flat. The event horizon is
again $r_{h} = 2 M$, and the surface gravity at the event horizon
is the same as that for a Schwarzschild metric, namely
\begin{equation}
\chi = 2 \pi T_{H} = \frac{1}{4 M}.
\end{equation}
This leads to the same classical thermodynamic properties
\cite{Ernst-properties} as for the case of Schwarzschild black
hole.
The massless scalar field equation has the form:
\begin{equation}\label{1}
\Box \Phi  \equiv \frac{1}{\sqrt{-g}} \left(g^{ \mu
\nu} \sqrt{-g}
\Phi,_{\mu}\right),_{ \nu} = 0
\end{equation}

\begin{figure}\label{figure8}
\caption{Effective potential for different values of the magnetic field: $B=0$ (bottom), $B=0.075$, $B=0.125$ (top)}
\resizebox{\linewidth}{!}{\includegraphics*{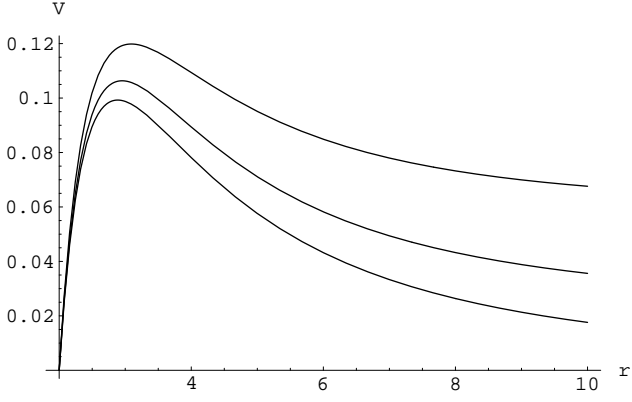}}
\end{figure}

In its general form, the Klein-Gordon equation does not allow
separation of radial and angular variables for the Ernst
background metric. Yet, even very strong magnetic fields in
centres of galaxies or in colliders, corresponds to $B \ll M$ in
our units, so that one can safely neglect terms higher than
$B^{2}$ in (\ref{1}). Indeed, in the expansion of $\Lambda^{4}$ in
powers of $B$, the next term after that proportional to $B^{2}
r^{2}$, is $\sim B^{4} r^{4}$ and, thereby, is very small in the
region near the black hole ($\Lambda^{4} \approx 1 + 4 B^{2} m^{2}
r^{2} + O(B^{4})$). The term $\sim B^{4} r^{4}$ is growing far
from black hole, and, moreover the potential in the asymptotically
far region is diverging, what creates a kind of confining by the
magnetic field of the Ernst solution. This happens because the
non-decaying magnetic field is assumed to exist everywhere in the
Universe. Therefore it is clear that in order to estimate a real
astrophysical situation, one needs to match the Ernst solution
with a Schwarzschild solution at some large $r$. Fortunately we do
not need to do this for the quasinormal mode problem: the
quasinormal spectrum of astrophysical interest is stipulated by
the behavior of the effective potential in some region {\it near
black hole}, while its behavior far from black hole is
insignificant
\cite{aetherPLB1_2}. In this way we take into consideration only
dominant correction due-to magnetic field to the effective
potential of the Schwarzschild black hole.
%Namely, neglecting in
%(\ref{1}) terms $\sim B^{4}$ and higher,
%after separation of
%angular variables, we reduce the wave equation (\ref{1}) to the
%Schroedinger wave equation
The exact Klein-Gordon equation for the angular part reads,
\begin{eqnarray}
\label{aux}
\frac{P_{sch}(\theta, \phi)\Phi}{r^2} + \frac{\Lambda^4 - 1}{r^2 \sin^2
  \theta}\partial_{\phi \phi}\Phi = 0,
\end{eqnarray}
with $P_{sch}(\theta, \phi)$ meaning the corresponding
pure-Schwarzschild part of the angular equation. Thus,
$P_{sch}(\theta, \phi)\Phi = - l(l+1)\Phi$ and neglecting in
(\ref{aux}) terms $\sim B^{4}$ and higher, after separation of the
angular variables, we reduce the wave equation (\ref{1}) to the
Schroedinger wave equation
\begin{equation}
\left(\frac{d^2}{dr_*^2}+\omega^2-V(r^*)\right)\Psi(r^*)=0,
\end{equation}
with the effective potential $V$:
\begin{equation}\label{2}
V(r)=f(r)\left(\frac{\ell(\ell+1)}{r^2}+\frac{2 M}{r^3}+ 4 B^{2}
m^{2} \right),
\end{equation}
where
$$f(r)=1-\frac{2 M}{r}, \quad dr^*=\frac{dr}{f(r)}$$
and $m$ is the azimuthal quantum number coming from the Killing
vector $\partial_{\phi}$ of the Ernst metric. We can see that the
effective potential (\ref{2}) coincides with the potential for the
{\it massive scalar field with the effective mass $\mu = 2 B m$ in
the Schwarzschild background}. Let us note also, that neglecting
higher terms in powers of $B^{2}$ we obtained the astro-physically
relevant wave equation which satisfies the quasinormal mode
boundary conditions: purely outgoing wave at spatial infinity and
pure in-going waves at the event horizon. The wave equation for
the exact Ernst metric would have diverging effective potential
(this can be seen for example from the full effective potential of
the particle moving in the equatorial plane around the Ernst black
hole \cite{Konoplya-Galtsov}) and would require the Dirichlet
confining boundary conditions.

In exactly the same way one can study the propagation of a scalar
field in the dilatonic Ernst background \cite{Agop:2005np}. The
metric has the form,
\begin{eqnarray}
\nonumber
d s^{2} = \Lambda^{\frac{2}{1+a^2}} \left(
\left(1- \frac{2 M}{r} \right) d t^{2}  -r^{2} d \theta^{2} \right. \\
\left. - \left(1- \frac{2 M}{r} \right)^{-1} dr^{2} \right)
 - \frac{r^{2} \sin^{2} \theta}{\Lambda^{\frac{2}{1+a^2}} } d
 \phi^{2},
\end{eqnarray}
where
\begin{eqnarray}
\Lambda = 1 + (1+a^2) B^2 r^2 \sin^2 \theta,
\end{eqnarray}
and $a$ is a constant factor which relates the dilaton and the
magnetic field as $\frac{\phi}{\ln \Lambda} = - \frac{a}{1+a^2}$.
The resulting effective potential is exactly the same as potential
(\ref{2}), so that the dilaton parameter $a$ cannot be seen  at
the dominant order in $B^2$ expansion.

The same occurs for the D-dimensional Ernst metric found by
Ortaggio \cite{Ortaggio:2004kr},
\begin{eqnarray}
\nonumber
ds^2 = \Lambda^{\frac{2}{D-3}}[-F(r)dt^2 + \{F(r)\}^{-1}dr^2 + r^2
  d\theta^2 \\
+ r^2 \cos^2 \theta d\Omega_{D-4}^{2}] + \Lambda^{-2} r^2 \sin^2 \theta d\Phi^2,
\end{eqnarray}
where $d\Omega_{D-4}^{2}$ is the line element of the
($D-4$)-sphere,
\begin{eqnarray}
d\Omega_{D-4}^{2} = d\Psi_1^2 + \prod_{a=1}^{D-5}\sin^2 \Psi_a d\Psi_{a+1}^{2},
\end{eqnarray}
the function $F(r)$ is
\begin{eqnarray}
F(r) = 1 - \frac{16 \pi M (D-2)^{-1}}{ r^{D-3}\Omega_{D-2}},
\end{eqnarray}
and $\Lambda$ is
\begin{eqnarray}
\Lambda = 1 + \frac{4(D-3)}{2D-4}B^2 r^2 \sin^2 \theta.
\end{eqnarray}
The potential for the scalar field propagation has the form
%of (\ref{aux1})
\begin{eqnarray}
\nonumber
V(r)=F(r)\left(\frac{\ell(\ell+D-3)}{r^2}+\frac{dF(r)}{dr}\frac{D-2}{2r}
+ 4 B^{2}m^{2} \right. \\
\left.+ \frac{F(r)(D-4)(D-2)}{4r^2} \right). \hspace{0.5cm}
\end{eqnarray}
If we again neglect the terms $\sim B^4$ and higher, the
quasinormal modes also have the same form for D-dimensional,
Dilatonic or ``pure Ernst'' geometry. Now we are in position to
use all the available data for the massive scalar quasinormal
modes in the Schwarzschild black hole. The quasinormal modes for
massive scalar fields were studied for the first time by Will and
Simone \cite{Will-Simone} and later in \cite{scalar} with the help
of the WKB method \cite{WKB}. The massive quasinormal modes are
characterized by the growing the damping time with the mass until
the appearance of the infinitely long lived modes called
quasi-resonances
\cite{quasiresonances}. In \cite{konoplya-zhidenkoPLB2005} it was shown
that when increasing the field mass,  the damped quasinormal modes
disappear "one by one" and a single corresponding quasi-resonance
appears instead, leaving all the remaining higher overtones
damped.

Note, that the wave equation with the obtained potential (\ref{2})
satisfies the quasinormal mode boundary condition at spatial
infinity, which in our particular case takes the form,
\begin{equation}\label{IB}\Psi(r^*)\sim C_+
e^{\imo\chi r^*}r^{(\imo M m^2/\chi)},
\quad (r,r^*\rightarrow+\infty)
\end{equation}
\begin{equation}
\chi = \sqrt{\omega^2-m^2}.
\end{equation}
(The sign of $\chi$ is to be chosen to stay in the same complex
surface quadrant as $\omega$.)
Within the Frobenius method we expand the wave function as follows
\cite{konoplya-zhidenkoPLB2005},
$$\Psi(r) = e^{\imo\chi r}r^{(2\imo M\chi+\imo M
m^2/\chi)} \times $$
\begin{equation}\label{FS}
\left(1-\frac{2M}{r}\right)^{-2\imo
M\omega}\sum_na_n\left(1-\frac{2M}{r}\right)^n,
\end{equation}

We obtained the quasinormal modes for different values of $B$,
$\ell$ and $m$. The results are summarized in the following table

\begin{table}
\caption{Quasinormal modes for Ernst black holes for different values of the magnetic field $B$, $M=1$, $D=4$.}
\vspace{3mm}
\begin{tabular}{|c|c|c|}
  \hline
  $B$ & $\ell =1$, $m=1$ & $\ell=2$, $m=1$ \\
  \hline
  0.005 & 0.292981 - 0.097633 i & 0.484433 - 0.096488 i \\
  0.025 & 0.294054 - 0.096988 i & 0.486804 - 0.095675 i \\
  0.050 & 0.297416 - 0.094957 i & 0.490764 - 0.094312 i \\
  0.075 & 0.303040 - 0.091521 i & 0.496327 - 0.092389 i \\
  0.100 & 0.321199 - 0.080040 i & 0.496327 - 0.092389 i \\
  0.125 & 0.333777 - 0.071658 i & 0.503512 - 0.089891 i \\
 % 0.150 & 0.348640 - 0.061174 i & 0.512346 - 0.086795 i \\
 % 0.175 & 0.365606 - 0.048285 i & 0.522862 - 0.083070 i \\
 % 0.200 & 0.384366 - 0.032754 i & 0.535100 - 0.078676 i \\
 % 0.225 & 0.404542 - 0.014468 i & 0.549107 - 0.073554 i \\
 % 0.250 & 0.483675 - 0.0096748 i & 0.564937 - 0.067625 i \\
\hline
\end{tabular}
\end{table}
\vspace{3mm}
The alternative time-domain description is based on the numerical
integration scheme described, for instance in \cite{time-domain}.
The obtained results show very good agreement with the accurate
numerical values of the table. For instance, for $B=0.05$
$m=\ell=1$, we have $0.295 - 0.096 i$ with the time domain
integration and $0.297416 - 0.094957 i$ with the Frobenius method.
The 6th order WKB formula for these values of parameters gives
$0.2974 - 0.0951 i$, what is very close to the accurate Frobenius
value. (The third order WKB formula certainly gives less accurate
value $0.2956 - 0.09527 i$). Let us note, that the WKB method is
expected to be much less accurate when applying to the massive
scalar field, because it implies pure exponential asymptotics at
both infinities and does not take into consideration the
sub-dominant asymptotic included in the pre-factor in \ref{IB}.
Nevertheless, when we are limited by relatively small values of
$B$, the 6th WKB formula is still a very good approximation.

The obtained numerical values of Table I is accurate as they are
based on the converging series expansion. The time domain
integration usually is less accurate.
\begin{figure}\label{figure1}
\caption{Evolution of perturbations for $B=0.05$, $M=1$, $m=1$, $\ell=1$, $D=4$.}
\resizebox{\linewidth}{!}{\includegraphics*{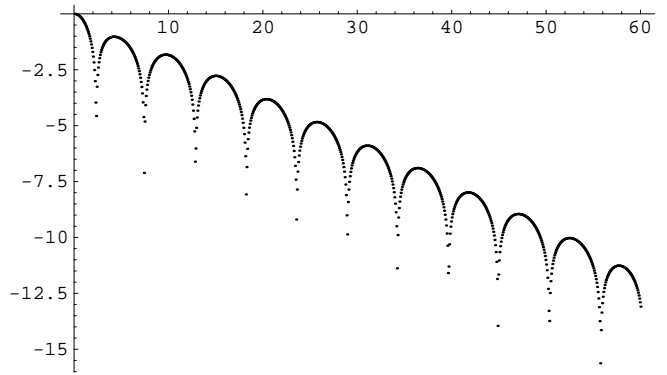}}
\end{figure}

From the available data we can see that the $Re \omega$ grows and
$Im \omega$ decreases when increasing the magnetic field $B$.
Therefore a black hole is a better oscillator in the presence of a
magnetic field, i.e. in this case the quality factor $Q \sim Re
\omega /Im \omega$ is considerably increased (see Fig. 2).
Let us note, that despite the azimuthal quantum number $m$ can be
large, it requires $\mid m \mid \leq \ell$, i.e. large $m$ will
require large $\ell$, so that the term $\sim 4 B^{2} m^{2}$ will
always be small in comparison with the centrifuge term $\sim \ell
(\ell +1) r^{-2}$. It means that the obtained effect is a small
correction to the Schwarzschild geometry.

It is essential that from the effective potential in Fig. 1, one
can see that for not very large $B$ the maximum of the potential
is slightly displaced from its Schwarzschild position $r= 3 M$. As
was shown in
\cite{aetherPLB1_2}, the quasinormal modes are stipulated by the
behavior of the effective potential near its maximum, because the
resonances of the scattering processes naturally depend on the
behavior of the potential near its pick. A good illustration of
this in \cite{aetherPLB1_2} shows that some potential given
numerically or by fit in such a way that it coincides with the
Regge-Wheeler potential near the black hole until $r = 5 M$ and
diverges at large $r$ has the same quasinormal modes as the
Schwarzschild black hole. This must persuade us that it is enough
to trust the Ernst solution until a few radius far from the black
hole and that one can safely neglect behavior in the further
region.

We have written down here only QNM values for $m=1$ and some first
$\ell$. Modes with higher $m$ can be easily obtained by
re-definition of the "mass term" $\mu_{eff} = 2 B m$.

\begin{figure}\label{figure2}
\caption{$Re \omega/Im \omega$ for different values of the magnetic field $B$: $M=1$, $m=1$, $\ell=1$, $D=4$.}
\resizebox{\linewidth}{!}{\includegraphics*{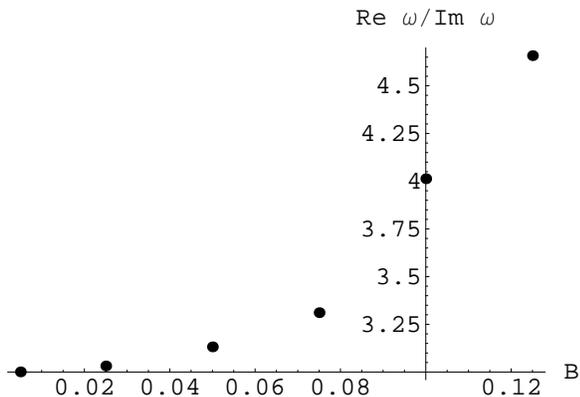}}
\end{figure}

Finally in Figures 3, 4, one can see the time domain profile for
the perturbations of higher dimensional generalization of the
Ernst black holes. The decay rate is again smaller in the presence
of the magnetic field, while the real oscillation frequencies are
larger, so that the quality factor is increased when the magnetic
field is increased. As there is quite complete numerical data on
quasinormal modes of massive scalar field for higher dimensional
Schwarzschild black holes in \cite{Zhidenko}, we showed here only
some time-domain profiles which give excellent agreement with the
numerical data of \cite{Zhidenko}.

\begin{figure*}
\resizebox{.5\linewidth}{!}{\includegraphics*{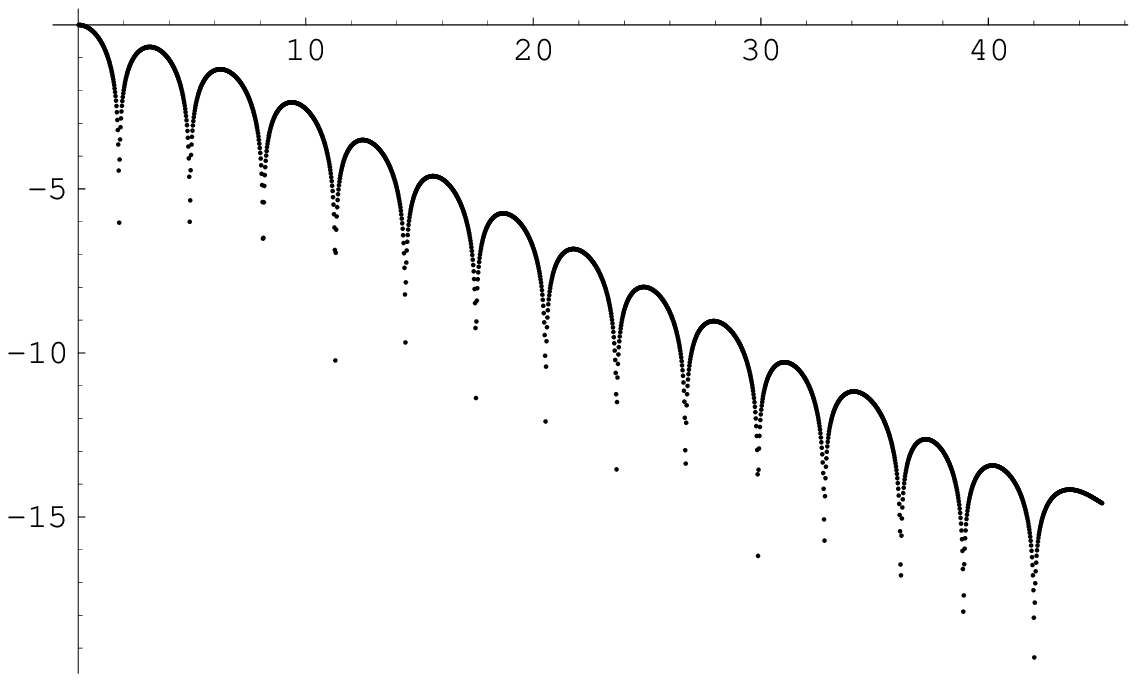}},\resizebox{.5\linewidth}{!}{\includegraphics*{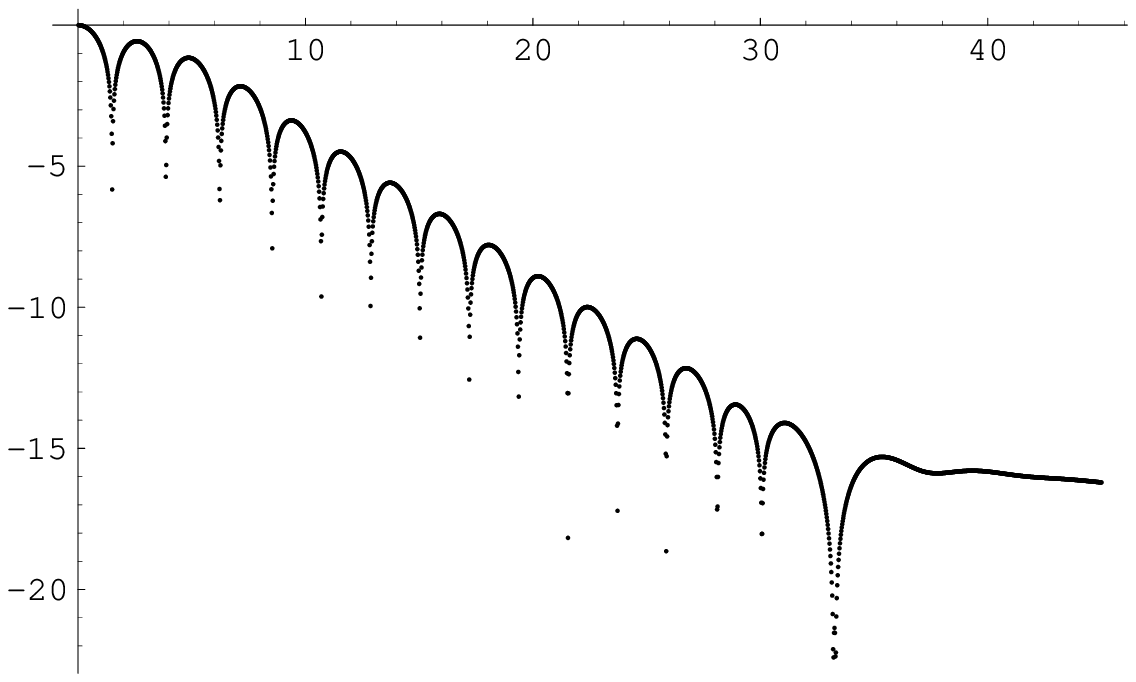}}
\caption{Time-domain profile for quasinormal ringing of the $\ell=m=1$ modes with $B=0.05$, $D=5$ (left), $D=6$ (right).}
\end{figure*}
\begin{figure*}
\resizebox{.5\linewidth}{!}{\includegraphics*{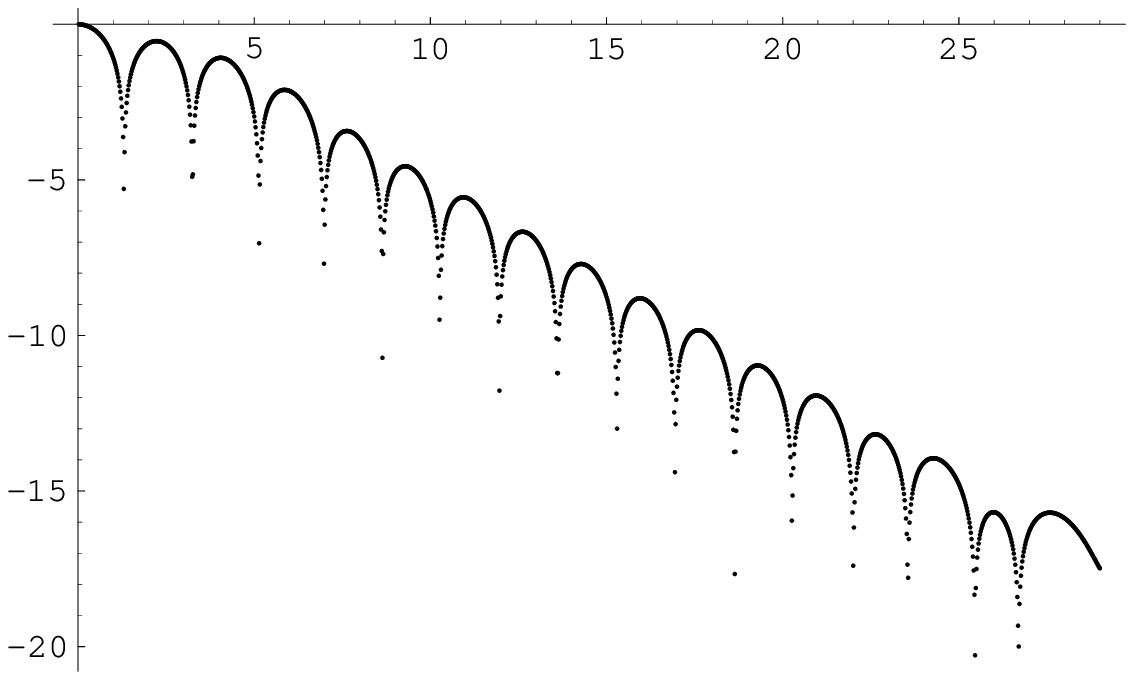}},\resizebox{.5\linewidth}{!}{\includegraphics*{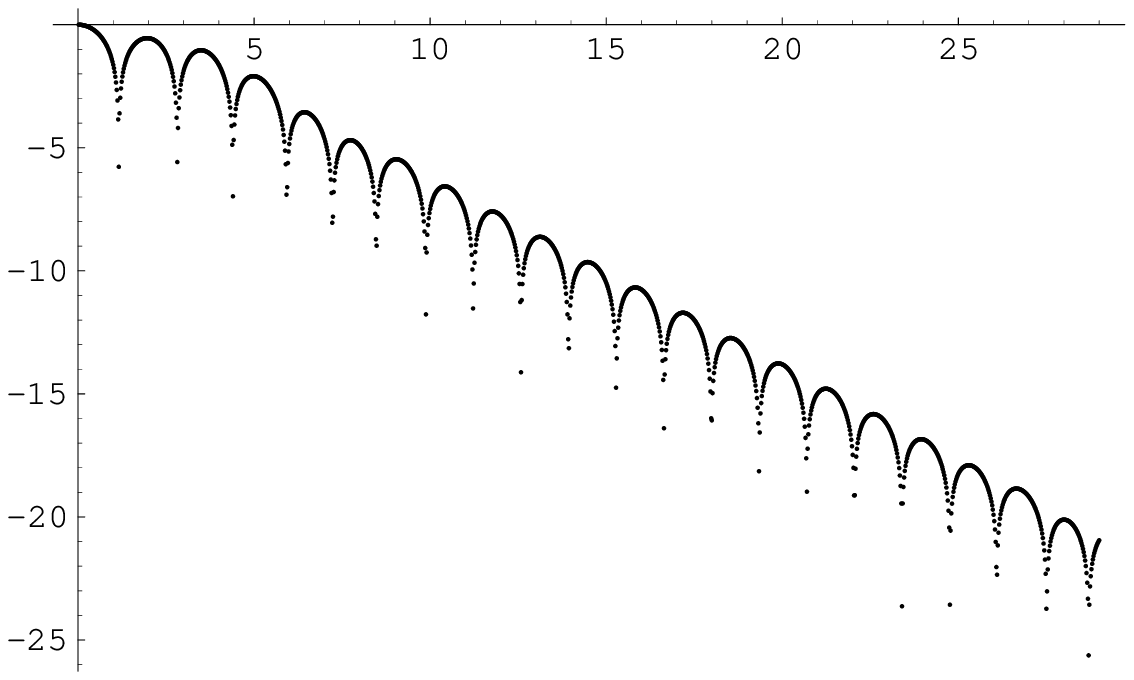}}
\caption{Time-domain profile for quasinormal ringing of the $\ell=m=1$ modes with $B=0.05$, $D=7$ (left), $D=8$ (right).}
\end{figure*}

\newpage
\vspace{1mm}
\textbf{Conclusion}
\vspace{1mm}

In the present paper we have found the quasinormal modes for the
Ernst black hole and its higher dimensional generalization. This
describes the influence of the strong magnetic field onto
characteristic quasinormal spectrum of black holes. In particular,
the real oscillation frequency grows when increasing the magnetic
field. The damping rate is decreasing, when the magnetic field is
increasing, so that magnetized black hole is characterized by
longer lived modes with higher oscillation frequencies, i.e. by
larger quality factor.

The obtained wave equation for the scalar field allows to find the
absorption cross-sections and quantum corrections to the entropy
of black holes due to scalar field in the vicinity of a strong
magnetic field \cite{Fontana-Konoplya}.

\begin{acknowledgments}
This work was supported by \emph{Funda\c{c}\~{a}o de Amparo
\`{a} Pesquisa do Estado de S\~{a}o Paulo (FAPESP)}. The authors
thank very useful discussions and advice of Elcio Abdalla and
Alexander Zhidenko.

%and
%\emph{Conselho Nacional de Desenvolvimento Cient\'ifico e Tecnol\'ogico (CNPq)},
%Brazil.

\end{acknowledgments}
%%%%%%%%%%%%%%%%%%%%%%%%%%%%%%%%%%%%%%%%%%%%%%%%%%%%%%%%%%%%%%%%%%%%%%%%%%%%%%%

\end{document}